\documentclass[%
superscriptaddress,
showpacs,
 amsmath,amssymb,
 aps,
prb,
twocolumn
]{revtex4-1}

\usepackage{lipsum}
\usepackage{graphicx}
\usepackage{dcolumn}
\usepackage{bm}
\usepackage{MnSymbol}
\usepackage{color}
\usepackage{dsfont}
\usepackage{hyperref}
\usepackage{float}
\usepackage[dvipsnames]{xcolor}

\begin{document}

\global\long\def\bra{\langle}
\global\long\def\brat#1{\langle#1|}
\global\long\def\ket{\rangle}
\global\long\def\kett#1{|#1\rangle}
\global\long\def\d{\partial}
\global\long\def\s#1{\mathcal{#1}}
\global\long\def\ex#1{\bra#1\ket}
\global\long\def\p#1{\left(#1\right)}
\global\long\def\ps#1{\left[#1\right]}
\global\long\def\qb#1#2{\langle#1|#2\rangle}
\global\long\def\mel#1#2#3{\langle#1|#2|#3\rangle}
\global\long\def\ple#1{\left(#1\right.}
\global\long\def\pre#1{\left.#1\right)}
\global\long\def\dag{\dagger}
\newcommand{\ks}[1]{\textcolor{red}{#1}}
\newcommand{\ch}[1]{\textcolor{blue}{#1}}
\newcommand{\mr}[1]{\textcolor{BurntOrange}{#1}}
\newcommand{\nl}[1]{\textcolor{magenta}{#1}}
\newcommand{\gr}[1]{\textcolor{cyan}{#1}}
\newcommand{\Ket}[1]{|#1\rangle}
\newcommand{\Bra}[1]{\langle#1|}

\preprint{APS/123-QED}

\title{Steady states of interacting Floquet insulators} 

\author{Karthik I. Seetharam}
\email{kseethar@caltech.edu}
\affiliation{Institute for Quantum Information and Matter, Caltech, Pasadena, California 91125, USA}
\author{Charles-Edouard Bardyn}
\email{charles.bardyn@unige.ch}
\affiliation{Department of Quantum Matter Physics, University of Geneva, 24 Quai Ernest-Ansermet, CH-1211 Geneva, Switzerland}

\author{Netanel H. Lindner}
\affiliation{Physics Department, Technion, 320003 Haifa, Israel}

\author{Mark S. Rudner}
\affiliation{Center for Quantum Devices and Niels Bohr International Academy,
Niels Bohr Institute, University of Copenhagen, 2100 Copenhagen, Denmark}

\author{Gil Refael}
\email{refael@caltech.edu}
\affiliation{Institute for Quantum Information and Matter, Caltech, Pasadena, California 91125, USA}

\date{\today}
\begin{abstract}
Floquet engineering offers tantalizing opportunities for controlling the dynamics of quantum many body systems and realizing new nonequilibrium phases of matter.  However, this approach faces a major challenge: generic interacting Floquet systems absorb energy from the drive, leading to uncontrolled heating which washes away the sought after behavior. How to achieve and control a non-trivial nonequilibrium steady state is therefore of crucial importance. In this work, we study the dynamics of  an interacting one-dimensional periodically-driven  electronic system  coupled to a phonon heat bath. Using the Floquet-Boltzmann equation (FBE) we show that the electronic populations of the Floquet eigenstates can be controlled by the dissipation. We find the regime in which the steady state features an insulatorlike filling of the Floquet bands, with a low density of additional excitations. Furthermore, we develop a simple rate equation model for the steady state excitation density that captures the behavior obtained from the numerical solution of the FBE over a wide range of parameters.  
\end{abstract}

\maketitle


\textit{Introduction \textendash }
Floquet engineering has emerged as an exciting tool for controlling the properties of quantum systems. 
A periodic drive, it was shown, could give rise to topological phases in graphene \cite{Oka2009,Kitagawa2011} as well as in trivial spin-orbit coupled semiconductors \cite{Lindner2011}. Subsequent work revealed a wealth of new 
phases without analogues in equilibrium\cite{KBRD, Jiang2011, Rudner2013, KhemaniPRL2016, Else_bauer_Nayak_PRL2016,Yao_2016,Lukin_2017,Monroe_2017,TitumPRX2016,PotterVishwanath2016,ElseNayak2016,KeyserlingkSondhi2016, RoyHarper2017,PoVishwanath2016,PoPotter2017,CarpentierGawedski2015}; these phases exhibit exotic features such as time-translation symmetry breaking\cite{Else_bauer_Nayak_PRL2016,Yao_2016,Lukin_2017,Monroe_2017}, topologically-protected chiral edge states in the presence of a completely localized bulk \cite{TitumPRX2016}, or fractionalized edges carrying a quantized flow of entropy \cite{PoVishwanath2016}. 

\begin{figure}[t]
\includegraphics[width=1\columnwidth]{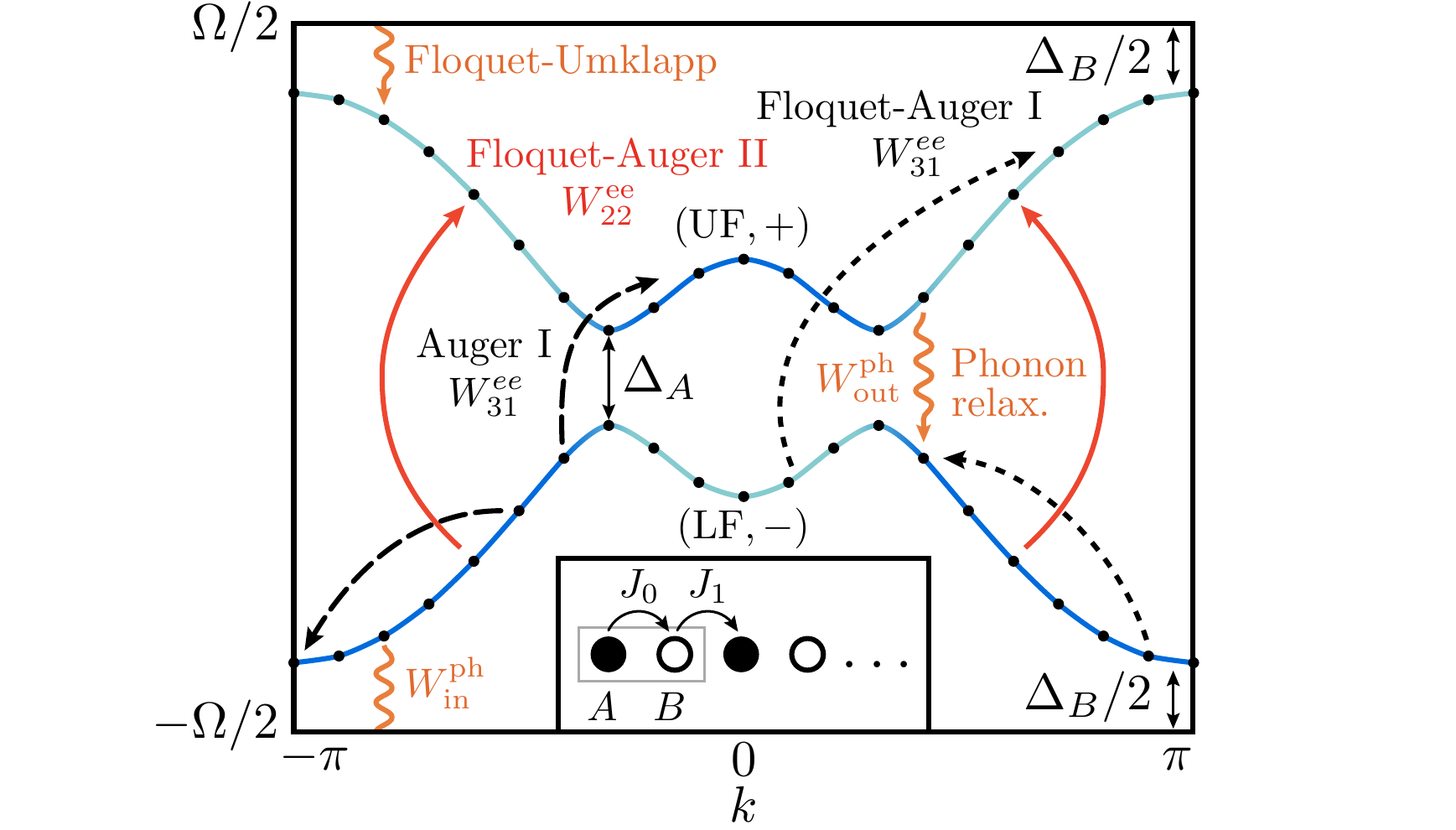}
\caption{Quasienergy band structure and interband scattering processes. Electron-electron interactions yield  three different types of interband processes: Auger, and Floquet-Auger (FA) of types I and II (see text) depicted by dashed, dotted, and solid lines, respectively. In the Floquet-Auger processes, the sums of quasienergies of the electrons in the initial and final states  differ by an integer multiple of the driving frequency, $\Omega$. Interband scattering resulting from electron-phonon interactions yields two important processes: (i) relaxation from the upper to the lower band via phonon emission, and (ii) excitation from the lower to the upper band. This process can occur even at zero temperature, as a  Floquet-Umklapp (FU) process, which involves phonon emission and absorption of $\Omega$ from the driving field.}
\label{fig:process}
\end{figure}

In many-body systems, Floquet engineering faces an important challenge due to electron-electron interactions. Interactions provide an efficient conduit for the system to absorb energy from the drive. In the absence of a bath, such energy absorption drives the system towards 
a maximum-entropy, infinite-temperature state \cite{Alessio_Rigol_2014_PRX,Lazarides_Moessner_2014_PRL,Lazarides_Moessner_2014_PRE,GenskeRosche2015,Bilitewski2014}.
Therefore, in order to assess the viability of Floquet engineering in electronic systems, it is crucial to determine the conditions under which a heat bath can stabilize a low-entropy steady state with certain key properties of interest.
In particular, in the context of trying to realize Floquet topological insulators, it is important that the steady state is well described in terms of electronic populations in the single-particle Floquet states.
Moreover, in order to observe the topological features of the system, we seek a population distribution corresponding to an insulator-like steady state.


Recently, several works have considered the steady states of non-interacting Floquet topological insulators in contact with external baths\cite{DehghaniMitra2014,DehghaniMitra2015,DehghaniMitra2015_2,DehghaniMitra2016,seetharam2015,IadecolaChamon2015-1,IadecolaChamon2015-2,ShiraiMiyashita2015,ShiraiMiyashita2016,Liu2015}.
These works showed that, under appropriate conditions on the driving and the system-bath coupling (such as phonon bandwidth\cite{seetharam2015, IadecolaChamon2015-1}, lead density of states\cite{seetharam2015, IadecolaChamon2015-2}, etc.), the topological features of the Floquet system may be observed through both the bulk Hall conductivity\cite{DehghaniMitra2015} and edge state transport\cite{EsinLindner2017}.
However, in the presence of interactions, it remains an open question whether the bath engineering strategies outlined in the works above are sufficient to control heating and stabilize the desired steady states.

In this work we consider the following question: can an insulatorlike filling of quasienergy bands be achieved in an interacting electronic system in which a periodic drive is used to induce a topological transition via a band inversion?
In this situation, the desired Floquet topological insulator (FTI) steady state is strikingly different from the ground state of the nondriven system: the FTI features a significant population inversion when viewed in terms of the valence and conduction bands of the host material.
Thus, in such a resonantly driven system, stabilizing the FTI steady state brings additional challenges compared to other protocols (e.g., based on high frequency driving).

To answer this question, here we consider a one-dimensional (1D) interacting, open, periodically-driven electronic system. 
We derive 
the Floquet-Boltzmann equation (FBE) for the electronic populations of the quasienergy states of the open interacting system~\cite{GenskeRosche2015,seetharam2015}. We numerically solve these equations for a system coupled to a bosonic bath of acoustic phonons, 
and show that, despite the interactions, the phononic bath still provides effective means for cooling the interacting driven system, even for experimentally realistic parameters. We develop a simple effective model for the Floquet band densities that captures the essence of all the Floquet scattering channels and that shows good numerical agreement with the exact FBE results for a large regime in parameter space. 

\textit{Microscopic model \textendash } To 0investigate dynamics of a periodically driven 1D electronic system, we employ a tight-binding model for spinless electrons with time-dependent hopping parameters and nearest-neighbor electron-electron interactions.  We consider a two-band model, with each unit cell of the lattice containing two sites (labeled $A$ and $B$, see inset of Fig.~\ref{fig:process}). The system's evolution is governed by the Hamiltonian $H =  H_0(t) + H_\mathrm{int}$, where the single-particle Hamiltonian 
\begin{equation}
\label{eq:quadratic}H_{0}(t) = \sum_{x}\Big([J_{0} +\delta J(t)]c_{x,A}^{\dag}c_{x,B} + J_{1}c_{x,B}^{\dag}c_{x+1,A}\Big) + \mathrm{h.c.}
\end{equation}
defines the system's band structure and driving, and
\begin{equation}
\label{eq:int}H_{\mathrm{int}}  =  V_0\sum_{x}\p{n_{x,A}n_{x,B}+n_{x,A}n_{x-1,B}}
\end{equation}
describes the nearest-neighbor interactions.
Here, $c^\dag_{x,A}$ and $c_{x,A}$ (likewise $c^\dag_{x,B}$ and $c_{x,B}$) denote the spinless electron
creation and annihilation operators on site $x$ of sublattice $A$ ($B$); the corresponding on-site densities are given by $n_{x,A}=c^\dag_{x,A}c_{x,A}$ and $n_{x,B}=c^\dag_{x,B}c_{x,B}$, respectively.  The intracell and intercell hopping parameters $J_0$ and $J_1$ as well as the interaction strength $V_0$ are taken to be positive and constant in time; throughout this work we take a modulation of the form $\delta J(t) = S\cos\Omega t$, where $\Omega$ is the drive (angular) frequency and $S$ is the driving strength. 

The single-particle Hamiltonian $H_0(t)$ in Eq.~(\ref{eq:quadratic}) is translationally invariant, and is therefore diagonal in crystal momentum. 
We introduce an index $\nu$ 
to label the bands of the system in the absence of driving, i.e., for $S = 0$.
In this basis, Eq.~(\ref{eq:quadratic}) takes the form $H_0(t) = \sum_{k\nu\nu'} c^\dagger_{k \nu}\left[E_{k}  \sigma^z_{\nu\nu'}  + \cos(\Omega t) (\bm{S}_k \cdot \bm{\sigma})_{\nu\nu'}\right]c_{k \nu'}$, where $E_k = |J_k|$ and $S_{k} = S(0,-\sin\theta_k,\cos\theta_k)$, with $J_k = J_0 + e^{i k a}J_1 \equiv |J_k| e^{i\theta_k}$.
Here, $\bm{\sigma} = (\sigma_x, \sigma_y, \sigma_z)$ is the vector of Pauli matrices, and $a$ is the lattice constant of the system.
We take the driving frequency $\Omega$ to be larger than the band gap of the nondriven system, $E_{\rm gap} = |J_0 - J_1|$, such that resonances are induced at crystal momenta $k_R$ satisfying $2E_{k_R} = \Omega$ (we set $\hbar=1$ throughout this work). 

In the presence of driving, the system is conveniently described in terms of its Floquet-Bloch band structure (see Fig.~\ref{fig:process}). 
We apply Floquet's theorem to find a complete basis of states $\Ket{\psi_{k \alpha}(t)} = e^{-i\mathcal{E}_{k \alpha}t}\Ket{\phi_{k \alpha}(t)}$ that satisfy Schr\"{o}dinger's equation with Hamiltonian $H_0(t)$, where $\Ket{\phi_{k \alpha}(t + T)} = \Ket{\phi_{k \alpha}(t)}$ is periodic with $T = 2\pi/\Omega$ and $\alpha = \pm$ labels the Floquet-Bloch bands with quasienergies $\mathcal{E}_{k \alpha}$.  
Importantly, the $T$-periodic function $\Ket{\phi_{k \alpha}(t)}$ can be expressed in terms of a discrete set of Fourier harmonics $\{\Ket{\phi_{k \alpha}^n}\}$, as $\Ket{\phi_{k \alpha}(t)} = \sum_n e^{-i n \Omega t}\Ket{\phi_{k \alpha}^n}$.
The structure of these harmonic coefficients plays an important role in determining the rates of the various scattering processes that will be considered below.

Equations (\ref{eq:quadratic}) and (\ref{eq:int}) prescribe the dynamics of the electronic system in isolation.  
In the presence of a periodic drive, the system's coupling to the environment plays a crucial role in determining its steady state. We therefore consider the electronic system's coupling to a bath of acoustic phonons.  
We take the system to be embedded in a three-dimensional (3D) medium which supports phonon modes, playing the role of the substrate supporting the 1D quantum wire.  
The phonon bath and electron-phonon coupling Hamiltonians are given by

\begin{eqnarray}
\label{eq:bosHam}H_\mathrm{b} & = & \sum_{\boldsymbol{q}}\omega_{\boldsymbol{q}}b_{\boldsymbol{q}}^{\dag}b_{\boldsymbol{q}},\\
H_{\mathrm{el-ph}} & = & \sum_{\boldsymbol{q}} \sum_{\substack{k\nu  \\ k'\nu'}}G_{\nu k}^{\nu'k'}\!\!(\boldsymbol{q}) \, c_{k'\nu'}^{\dag}c_{k \nu}(b_{\boldsymbol{q}}+b_{-\boldsymbol{q}}^{\dag}).
\label{eq:elbosHam}
\end{eqnarray}
Here, $\boldsymbol{q}=(q,\bm{q}_{\perp})$ is the phonon momentum (with components $q$ parallel to the 1D electronic system, and $\bm{q}_{\perp}$ in the transverse direction), and $\omega_{\boldsymbol{q}} = C|\bm{q}|$ defines the phonon spectrum, taken to be linear and isotropic with speed of sound $C$, up to a frequency cutoff $\Omega_D$. 
The electron-phonon interaction amplitude $G_{\nu k}^{\nu'k'}\!\!(\boldsymbol{q})$ corresponds to an electronic transition $\nu k \rightarrow \nu'k'$ via absorption of a phonon with momentum $\bm{q}$ (or emission with $-\bm{q}$); this amplitude is proportional to $\sum_l \delta(k'-k-q+2\pi l/a)$, with $l$ ranging over all integers, ensuring crystal-momentum conservation along the direction of the electronic system. 
For simplicity, in this work we choose the matrix elements multiplying the momentum delta function in the phonon scattering amplitude to be $G_0 \sigma^3_{\nu\nu'}$; i.e., the electron-phonon coupling conserves the band index of the non-driven system. 
The qualitative features of our results do not depend on the exact form of the electron-phonon coupling.
The Debye cutoff frequency $\Omega_D$ is an important parameter of the model, which we use to control the types of possible scattering processes (see below). 

We seek the steady states of the interacting driven system coupled to the bosonic (phonon) bath described by Eqs.~(\ref{eq:bosHam}) and (\ref{eq:elbosHam}). We define the population of the single-particle Floquet state $k\alpha$ 
 as $F_{k\alpha}(t) = \langle f^\dagger_{k\alpha}(t) f_{k\alpha}(t)\rangle$, where the operator {$f^\dagger_{k\alpha}(t) = \sum_{\nu,n}e^{-i (\mathcal{E}_{k\alpha} + n\Omega)t}\langle k \nu | \phi^n_{k\alpha}\rangle c^\dagger_{k\nu}$ creates an electron in the Floquet state $\Ket{\psi_{k\alpha}}$ at time $t$. We focus on the regime where scattering rates in the steady state are small compared with the gaps between Floquet-Bloch bands, translation invariance is maintained, and strong multi-particle correlations (e.g., excitons) are absent. In this regime, the steady state is well represented in terms of the populations $F_{k\alpha}(t)$ of the single-particle Floquet states.
We use the Floquet-Boltzmann equation (FBE) \cite{GenskeRosche2015,Bilitewski2014,seetharam2015} to evolve these populations:
\begin{equation}
\label{eq:FBE_main} \dot{F}_{k\alpha} = I^{\rm ph}_{k\alpha}(\{F\}) + I^{\rm ee}_{k\alpha}(\{F\}),
\end{equation}
where $I^{\rm ph}_{k\alpha}$ and $I^{\rm ee}_{k\alpha}$ are the collision integrals that capture the net rates of electron scattering into Floquet state $k\alpha$ due to electron-phonon and electron-electron interactions, Eqs.~(\ref{eq:elbosHam}) and (\ref{eq:int}), respectively.
Explicit expressions for these collision integrals and the Fermi's golden rule transition rates inside them are given in Appendix \ref{sec:Derivation}.

\textit{Simple model for population kinetics \textendash} Before examining the numerical solution of the full FBE, we first develop and discuss a simple effective model that captures the basic qualitative features of the steady states of Eq.~(\ref{eq:FBE_main}). 
Specifically, we focus on the interplay between electron-electron and electron-phonon scattering in determining the 
net populations of the two Floquet-Bloch bands,
\begin{equation}
n_{\alpha}  =  \frac{1}{N}\sum_{k}F_{k\alpha}, 
\label{eq:density}
\end{equation}
where $\alpha=-,+$ denotes the lower/upper Floquet (LF/UF) bands, respectively (see Fig.~\ref{fig:process}), and $N$ is the number of unit cells in the system. 
At half filling, which is our focus in this work, the number of excitations in the upper Floquet band is equal to the number of holes in the lower Floquet band; this implies 
$n_+ = 1-n_{-}\equiv n$.

Due to the periodicity of quasienergy, the designation of ``upper" and ``lower" Floquet bands amounts to a gauge choice. However, the rates of dissipative processes are sensitive to the characters of the Floquet band wave functions (valence-band-like or conduction-band-like), and provide a natural orientation for the bands (see, e.g., Refs.~\onlinecite{EsinLindner2017,seetharam2015}). 
Our choice follows this natural orientation, picked in anticipation of the results below.

We construct the model by characterizing the rates of all possible inter-Floquet-band transitions facilitated by electron-phonon scattering and electron-electron interactions. 
The rates of the various scattering processes depend on incoming and outgoing crystal momentum and band indices, as well as the full distribution of Floquet state populations, $\{F_{k\alpha}\}$, see Eq.~(\ref{eq:FBE_main}).
Therefore, the evolution of the excitation density $n$ generally cannot be written as a function of $n$ alone.
As a crude approximation, a closed dynamical equation for $n$ can be obtained by making a ``uniform'' approximation on the FBE, replacing all $k$-dependent rates by their band-averaged values (see Appendix~\ref{sec:Derivation}).
Crucially, this model retains the essential structure of phase-space restrictions on different classes of processes, which we describe in detail below.
Comparing to numerical simulations of the full FBE, we will show that the simple model captures and provides insight into the qualitative dependence of the steady-state excitation density on the fundamental parameters of the system. 

Consider first the possible electron-phonon scattering processes. 
Phonon-mediated transitions {\it out} of the UF band (and into the LF band) require an excited particle in the UF band to scatter into a hole in the LF band. 
This requirement constrains the phase space for such processes, which thus provides a sink for density in the UF band with rate  $W^{\rm ph}_{\rm out} n_+ (1-n_-)=W^{\rm ph}_{\rm out} n^2$. We refer to processes that reduce the density of excitations as ``cooling'' processes. 
Similarly, phonon-mediated transitions from the LF band {\it into} the UF band require a particle in the LF band to scatter into an empty state in the UF band.
Such processes provide a source for the excited population, with rate $W^{\rm ph}_{\rm in} (1-n_+) n_-=W^{\rm ph}_{\rm in} (1-n)^2$. We refer to processes that increase the density of excitations as ``heating'' processes. 

Importantly, the competition between phonon-mediated ``heating'' and ``cooling'' processes, captured by the  rates $W^{\rm ph}_{\rm in}$ and $W^{\rm ph}_{\rm out}$, depends on the driving strength and frequency, as well as the bandwidth of the phonon bath, $\Omega_D$. 
We consider the case where the phonon bandwidth is larger than the resonance-induced Floquet gap centered at quasi-energy $\mathcal{E}=0$, denoted by $\Delta_A$ in Fig.~\ref{fig:process}. 
Under this condition, the sink rate $W^{\rm ph}_{\rm out}$ in Eq.~(\ref{eq:effDyn}) is nonzero; excited particles in the UF band can scatter into available holes in the LF band, while emitting a phonon to conserve quasienergy. 
In contrast, at zero temperature (and assuming $\Omega_D<E_{\rm gap}$), scattering processes contributing to the bare rate $W^{\rm ph}_{\rm in}$ in the source term are always of ``Floquet-Umklapp'' type: the scattered electron's quasienergy in the final state differs from its initial value by $\Omega-\omega_\mathbf{q}$, where $\omega_\mathbf{q}$ is the energy of the emitted phonon. For $\Omega_D < E_{\rm gap}$ and/or an electron-phonon coupling that is diagonal in the original band indices, we find that the rate $W^{\rm ph}_{\rm in}$ is suppressed in comparison to $W^{\rm ph}_{\rm out}$ by a factor of $(S/\Omega)^4$ (where $S$ is the drive strength, and $\Omega$ is its frequency), see App.~\ref{sec:Derivation}. 
Thus for weak driving, $(S/\Omega) \ll 1$, heating due to electron-phonon scattering is naturally a weak effect (see Fig.~\ref{fig:scalingEl} for more details).

Electron-electron interactions may give rise to two types of ``Auger'' processes that can change the populations in the two Floquet bands: (I) two particles in the same Floquet band may scatter to a final state which has one particle in each of the Floquet bands, and (II) two particles in the same Floquet band may simultaneously scatter to the opposite Floquet band. Examples of these processes are depicted in Fig.~\ref{fig:process} (see also Fig.~\ref{fig:scalingEl}). 

Electron-electron scattering conserves total crystal momentum and quasi-energy.  Similar to conservation of crystal momentum, conservation of quasienergy can either be ``direct,'' with the sum of initial and final single particle quasienergies being equal, or ``Umklapp''-like, where the sum of single particle quasienergies in the final state differs from its initial value by $\Omega$. Processes of type (I) can be either direct or Floquet-Umklapp-like; we label such processes ``Auger I'' and ``Floquet-Auger I,'' respectively. Processes of type (II), which we label ``Floquet-Auger II,'' are necessarily of the Umklapp type. For weak driving, the rates of these Floquet-Umklapp processes are suppressed by a factor $(S/\Omega)^2$ (for a specific Floquet-Umklapp process, the suppression can be even stronger). 



We now characterize the rates for electron-electron scattering processes, taking into account the phase-space requirements for the corresponding transitions.
Processes of type I require two particles in the initial band to scatter into two empty states, one in each band.  
If the two particles are initially in the LF band, we obtain a source term for the excitation density (a ``heating'' process) with rate $W^{\rm ee}_{31}n_-^2(1-n_+)(1-n_-)=W^{\rm ee}_{31}(1-n)^3n$. 
Note that this rate includes the contributions of both Auger-I and Floquet-Auger I processes.
If both particles are initially in the UF band, we obtain a sink term for the density of excitations with a rate of $W^{\rm ee}_{31}n_+^2(1-n_+)(1-n_-)=W^{\rm ee}_{31}n^3(1-n)$. Due to particle-hole symmetry, the same bare rate $W^{\rm ee}_{31}$ appears for both the source and sink terms. 


Using similar considerations, we find that processes of type II contribute a source term for $n$ with rate $W_{22}^{\rm ee} n_-^2(1-n_+)^2=W_{22}^{\rm ee}(1-n)^4$, and a sink term with rate $W_{22}^{\rm ee} n_+^2(1-n_-)^2=W_{22}^{\rm ee}n^4$. 
In the primary regime of interest the excitation density will be small.  Therefore, the {\it sink terms} arising from electron-electron scattering will be suppressed (relative to the source terms), as they involve higher powers of $n$.

Combining all source and sink terms, the rate of change of the excitation density $n$ is approximately given by 
\begin{eqnarray}
\dot{n} & = & W^{\rm ph}_{\rm in}(1-n)^{2}-W^{\rm ph}_{\rm out} n^{2} +W_{31}^{\rm ee}[n(1-n)^{3}-n^{3}(1-n)] \nonumber \\
 &  & +W_{22}^{\rm ee}[(1-n)^{4}-n^{4}]. \label{eq:effDyn}
\end{eqnarray}
We obtain the steady-state population of the UF band 
by solving $\dot{n}=0$. 
This condition yields a cubic equation for the steady-state excitation density, which is supplemented by the condition $0 \le n \le 1$. While such a relation in principle admits for multistability, we find only a single physical solution in all regimes studied.
 In App.~\ref{sec:Lead} we present a generalization of Eq.~(\ref{eq:effDyn}) which incorporates the role of a fermionic reservoir.

Although Eq.~(\ref{eq:effDyn}) can be solved exactly using the general solution for the roots of a cubic polynomial, it is instructive to examine the behavior perturbatively around specific limits of interest. In the absence of phonons, $W^{\rm ph}_{\rm in}=W^{\rm ph}_{\rm out}=0$, interactions drive the system toward a high-entropy state with $n^*=1/2$. In the more general scenario, the phonon bath can extract entropy and energy from the system, yielding a non-trivial steady state. 

A nontrivial steady state with a Floquet-band-insulator-like distribution is obtained when the heating rates due to electron-phonon and electron-electron interactions are small compared with the rate of relaxation by the phonon bath.
To characterize this regime, it is useful to define the dimensionless quantities $\kappa_{31}=W_{31}^{\rm ee}/W^{\rm ph}_{\rm out}$, 
$\kappa_{22}=W_{22}^{\rm ee}/W^{\rm ph}_{\rm out}$ and $\kappa_{\rm ph}=W_{\rm in}^{\rm ph}/W^{\rm ph}_{\rm out}$. As explained above, we expect $\kappa_{\rm ph} \ll 1$.
For weak interactions, we may also have $\kappa_{22}, \kappa_{31} \ll 1$.
Within this limit, the excitation density in the steady state will be small, $n\ll1$.  To lowest order in $n$,  the heating rate in Eq.~\eqref{eq:effDyn} arising from electron-electron scattering is $W_{22}^{\rm ee}$. Therefore, if  $\kappa_{\rm 22} \gg \kappa_{\rm ph}$, electron-electron scattering provides the main source of heating and we find $n^* \sim \sqrt{\kappa_{22}}$. When electron-phonon scattering dominates the heating rate, $\kappa_{\rm ph} \gg \kappa_{\rm 22}$, we expect $n^* \sim \sqrt{\kappa_{\rm ph}}$.

\begin{figure}
\includegraphics[width=\columnwidth]{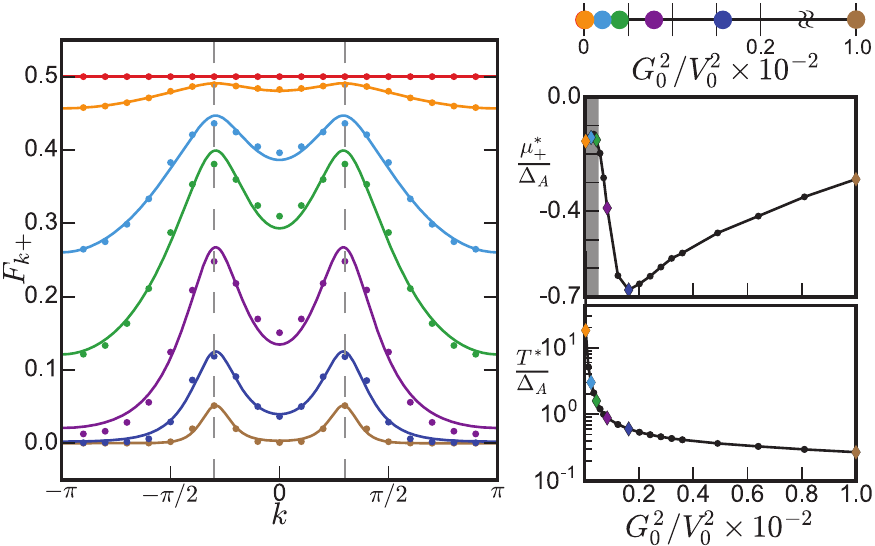}
\caption{Left: Steady-state populations in the UF band, $F_{k+}$, for several values of the effective cooling strength $G_0^2/V_0^2$.
Results are obtained from the FBE, Eqs.~(\ref{eq:FBE_main}) and (\ref{eq:FBE}), with phonon bandwidth $\Omega_D/\Delta_A=2.2$ and phonon temperature $T_{\rm ph} = \Delta_A/10$. 
Dashed lines indicate the crystal momentum values where the UF band minima are located.
For low values of $G_0^2/V_0^2$, the steady state is ``hot,'' with nearly uniform occupation $F_{k+} \approx 0.5$ for all $k$.
For large values of $G_0^2/V_0^2$, the steady state is ``cold,'' and features a low density of excitations concentrated around the minima of the UF band.
Solid lines show fits to a Floquet-Fermi-Dirac distribution with effective chemical potential $\mu^*_+$ (with respect to $\s E =0$), and temperature $T^*$, taken as free parameters.
Right: extracted values of $\mu^*_+$ and $T^*$ {\it vs.}~$G_0^2/V_0^2$.
When $\mu^*_+ \neq 0$, the steady state is described by a ``double'' Floquet-Fermi-Dirac distribution, with separate chemical potentials for electrons and holes in the UF and LF bands, respectively. The shaded region in upper panel denotes a regime where the fits are sensitive only to the value of $T^*$ (and are insensitive to the value of $\mu^*_+$). 
}
\label{fig:dist}
\end{figure} 

\textit{Results \textendash } We now discuss numerical results for the solution of the full Floquet Boltzmann equation, Eq.~(\ref{eq:FBE_main}), and their comparison with the predictions of the simple model described above.
In Fig.~\ref{fig:dist} we show the full momentum-resolved steady-state populations in the UF band, for several ratios of the electron-phonon ($G_0$) and electron-electron ($V_0$) coupling strengths [see Eq.~(\ref{eq:int}) and text below Eq.~(\ref{eq:elbosHam})].

To start from a conceptually simple case, in Fig.~\ref{fig:dist} we take a restricted phonon bandwidth $\Omega_D < \Delta_B$ (see Fig.~\ref{fig:process}), which ensures that phonon-mediated Floquet-Umklapp processes are energetically forbidden.
Under this condition, the only source terms for excitation density (i.e., ``heating processes'') are electron-electron-mediated Floquet-Umklapp processes and thermally-activated phonon absorption.
The rates of the latter are suppressed by a factor $e^{-\Delta_A/T_{\rm ph}} \approx 5 \times 10^{-5}$ for $T_{\rm ph} = \Delta_A/10$, as used in the simulations.
To a very good approximation, in this regime, $G_0$ controls cooling and $V_0$ directly controls heating. 

As a function of the ratio $G_0^2/V_0^2$ 
we observe a clear transition from a ``hot'' state with nearly uniform populations,  $F_{k\pm}\approx 0.5$ for all $k$, to a ``cold'' state in which the LF (UF) band is nearly completely filled (empty). The ``cold'' state hosts a small density of excitations near the band extrema around $\mathcal{E}=0$. We fit the populations $F_{k\pm}$ using two separate Floquet-Fermi-Dirac distributions, with independent chemical potentials $\mu^*_+$ and $\mu^*_-$ for electrons and holes in the upper and lower Floquet bands, respectively. By particle-hole symmetry, $\mu^*_{-} = -\mu^*_+$. The fits are shown as solid lines in Fig.~\ref{fig:dist}. The effective temperature $T^*$ and chemical potential $\mu^*_+$ extracted from these fits are shown in the upper and lower panels on the right of Fig.~\ref{fig:dist}. Note that without phonon-mediated Floquet-Umklapp processes and in the $V_0=0$ limit, the ``global'' Floquet-Gibbs state with populations $F_{k\alpha}=(e^{\s{E}_{k\alpha}/T_\mathrm{ph}}+1)^{-1}$, i.e., with $\mu^*_{-} = \mu_+^*=0$, is an exact solution to the FBE (see Appendix~\ref{sec:Derivation} and Refs.~\onlinecite{GalitskiiElesin1970,seetharam2015,ShiraiMiyashita2015,ShiraiMiyashita2016,Liu2015}).
In particular, in this limit and for $T_\mathrm{ph}=0$, the steady-state is an ideal Floquet insulator state with $F_{k-}=1$ and $F_{k+} = 0$ for all $k$.

\begin{figure}
\includegraphics[width=\columnwidth]{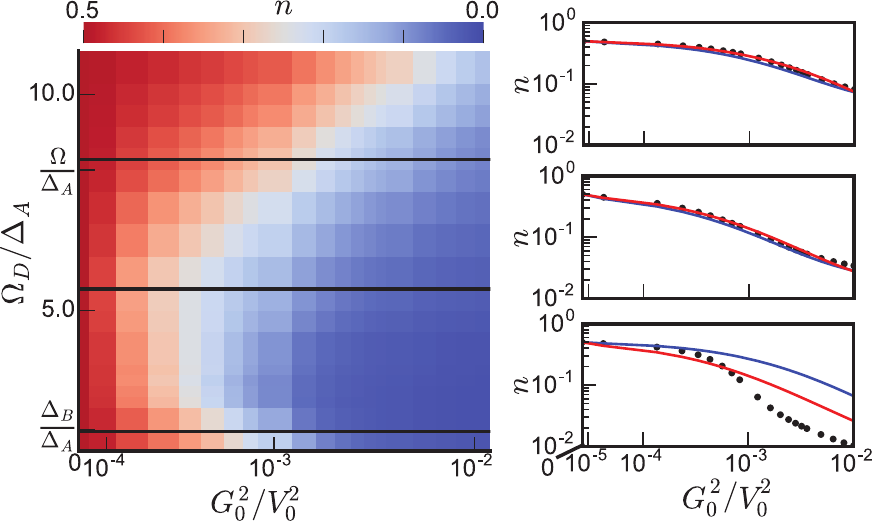}
\caption{Left: Excitation density $n = n_+$, Eq.~(\ref{eq:density}), as a function of the (normalized) phonon bandwidth $\Omega_D/\Delta_A$ and $G_0^2/V_0^2$. For large $G_0^2/V_0^2$, the phonon bath effectively cools the system, and the steady-state excitation density is low (blue color).
The cutoff $\Omega_D$ controls the phase space for electron-phonon scattering; the cooling effect of the phonon bath is strongest for intermediate values of $\Omega_D$ where many relaxation processes are allowed, and heating due to phonon-mediated Floquet-Umklapp processes is relatively suppressed. Note that $\Delta_B/\Delta_A=2.25$ and $\Omega/\Delta_A=8.25$. Right (from top to bottom): Line cuts at $\Omega_D/\Delta_A=8.5,5.5,2.2$. Blue lines show results from the effective model (Eq.~\ref{eq:effDyn}) using rates computed by direct application of the uniform approximation. Red lines indicate the results of the effective model with fitted parameters (see main text). 
For $\Omega_D/\Delta_A=8.5,5.5$, the average rates are quite close to the best fit curves and also give a good approximation to the exact FBE data. 
For $\Omega_D/\Delta_A=2.2$, the scattering phase space is highly restricted and the simple model does not provide a good description of the FBE results.
}
\label{fig:effModel}
\end{figure} 

Going beyond the restricted scenario of Fig.~\ref{fig:dist}, we now examine how the steady state is affected by phonon-mediated Floquet-Umklapp processes when the phonon bandwidth $\Omega_D$ is increased.
The excitation density $n$ [Eq.~(\ref{eq:density})] as a function of $\Omega_D$ and $G_0^2/V_0^2$ is shown in Fig.~\ref{fig:effModel}. 
Although increasing $G_0$ increases the rates of both phonon-mediated cooling and heating processes, the blue color on the right side of Fig.~\ref{fig:effModel} indicates that increasing $G_0$ (for fixed $V_0$) has the overall effect of decreasing the excitation density. This can be understood by recalling that for $\Omega_D<E_{\rm gap}$, phonon-mediated Floquet-Umklapp transition rates are suppressed with respect to direct transitions by a factor of $(S/\Omega)^4$. 

The excitation density exhibits a non-monotonic dependence on $\Omega_D$, which we interpret as follows: 
In the regime $\Delta_A<\Omega_D<\Delta_B$, as considered in Fig.~\ref{fig:dist}, phonon-mediated interband relaxation (cooling) is possible, but the corresponding FU processes are forbidden. 
However, for low values of $\Omega_D$ the scattering phase space is restricted and cooling is inefficient.
As $\Omega_D$ is increased, the phase space for electron-phonon scattering increases and the bath is able to cool the system more effectively.
When $\Omega_D > \Delta_B$, 
phonon-mediated FU processes are allowed and compete with the cooling effect of the bath.
This competition leads to an optimal value  $\Omega^{\rm opt}_D>\Delta_B$ where the excitation density is minimized for a given value of $G_0^2/V_0^2$.



We now compare the results for the numerical solution of the FBE to the predictions of the simple effective model described above (right three panels of Fig.~\ref{fig:effModel}). We consider two approaches for determining the effective rate parameters in Eq.~(\ref{eq:effDyn}). In the first approach, we average the bare rates over momentum as per the uniform approximation in Eq.~(\ref{eq:avgRatesPH}) and Eq.~(\ref{eq:avgRatesEE}), and use them to predict the steady-state ($\dot{n}=0$) for each case of $\Omega_D$ and $G_0^2/V_0^2$. The second approach builds on the first. For a given $\Omega_D$, the average rates $W_{\mathrm{in}}^{\mathrm{ph}},W_{\mathrm{out}}^{\mathrm{ph}},W_{31}^{\mathrm{ee}}$, and $W_{22}^{\mathrm{ee}}$ form four separate functions of $G_0^2/V_0^2$. 
We introduce a scaling prefactor each of these functions, which we use as fitting parameters. (Note that a global rescaling of all four functions leaves the steady state invariant; hence there are three independent fitting parameters.)
These three parameters are fitted using the method of least squares for the difference between the predicted densities from the effective model and the exact densities computed from the FBE (taken over all values of $G_0^2/V_0^2$).

The simple model in Eq.~(\ref{eq:effDyn}) is based on a ``uniform'' approximation, in which the crystal momentum dependencies of the transition rates and populations are ignored. As such, we expect the simple model to work well in the ``hot'' regime where the distribution approaches a uniform, infinite-temperature-like form.
Interestingly, when the phonon bandwidth is large, $\Delta_B<\Omega_D<E_{\rm gap}$,  we observe good agreement between the effective model and the full FBE even well outside the hot regime, where the total excitation density becomes small (see upper two line cuts in Fig.~\ref{fig:effModel}). Furthermore, in this regime, we see that the two methods for determining the effective rates in Eq.~(\ref{eq:effDyn}) give very similar results. 
 For lower values of $\Omega_D$ (lowest panel, with $\Delta_A < \Omega_D < \Delta_B$), the phase space for electron-phonon scattering becomes highly restricted and we observe significant deviations between the solution of the FBE and the simplified model.

\emph{Discussion \textendash}
Our motivation in this work was to study the applicability of Floquet band engineering in the presence of electron-electron interactions. In particular, we were interested in the situation occurring in Floquet topological insulators, where a resonant drive induces a band inversion in the Floquet spectrum.  We find the regime where cooling by the phonon bath effectively counters the heating mediated by the interactions, thereby stabilizing an insulator-like steady state with a small density of excitations. 

To identify the experimentally-relevant regime, we now relate our model parameters to typical time scales observed in driven semiconductors. The shortest timescale is associated with elastic electron-electron interactions, $\tau^{\rm elastic}_{\rm ee}\sim 10-100$ fs}, while the cooling timescale due to electron-phonon scattering is on the order of $\tau_{\rm ph} \sim  0.1-1$ ps \cite{SundaramMazur2002}. As discussed above, in the low-excitation-density regime, Floquet-Auger II processes dominate the heating rate. These processes are of Floquet-Umklapp type, and we thus estimate the associated time scale to be $\tau^{\rm FU}_{\rm ee}=(W_{22}^{\rm ee})^{-1}\sim(S/\Omega)^{-2}\tau^{\rm elastic}_{\rm ee}$. Therefore, a rough estimate for the dimensionless parameter controlling the excitation density is $\kappa_{22}=(S/\Omega)^2\tau_{\rm ph}/\tau^{\rm elastic}_{\rm ee}$. For $(S/\Omega)\lesssim 0.1$, a regime of low excitation density can be reached. 

To simplify the analysis in this work, we did not consider electron-hole radiative recombination processes, which also contribute to heating \cite{seetharam2015}. These processes can be straightforwardly incorporated to the model. At the level of the effective model in Eq.~(\ref{eq:effDyn}), recombination processes only renormalize the parameters $W^\mathrm{ph}_\mathrm{out}$, $W^\mathrm{ph}_\mathrm{in}$. The radiative recombination time scale is on the order of $\tau_{\rm r}\sim 0.1\ {\rm ns} \gg \tau^{\rm elastic}_{\rm ee} $. Thus, the contribution of radiative recombination to heating will be dominant only for $(S/\Omega)^2\ll 1$.

A further simplification in our model was the choice of band structure parameters to allow only a single-photon resonance, see Fig.~\ref{fig:process}.  Floquet gaps resulting from an $n^{\rm th}$-order resonance  would be suppressed by a factor of $(S/\Omega)^n$. Thus, in many experimental realizations, we expect these gaps to be smaller than the scattering rates in the steady state. Therefore, the primary role of the higher-order resonances would be to add additional heating channels, whose rates would be suppressed by corresponding powers of $(S/\Omega)$. Their effect would be subdominant, and would not change our results qualitatively. The effect of higher-order resonances for strong driving is an interesting direction for future work.

Our demonstration that the populations of the Floquet bands can be controlled in the presence of electron-electron interactions leaves many directions for future research: In the regime of low excitation density, an important goal is to find experimental probes for extracting the topological properties of the Floquet band structure. For higher excitation densities, we have shown that it is possible to reconstruct the results of the full FBE with a simple, nonlinear rate equation, Eq.~(\ref{eq:effDyn}). 
The effective model opens an interesting route for exploring the interplay between nonlinear phenomena such as bistability and hysteresis with the physics of Floquet-engineered band structures.



\acknowledgements We thank L. Glazman, Yuval Baum, Evert van Nieuwenburg, Justin Wilson, Michael Buchhold, and Min-Feng Tu, for useful discussions. N.L. acknowledges support from the European Research Council (ERC) under the European Union Horizon 2020 Research and Innovation Programme (Grant Agreement No. 639172), from the People Programme (Marie Curie Actions) of the European Union’s Seventh Framework 546
Programme (FP7/2007–2013), under REA Grant Agreement 
No. 631696, and from the Israeli Center of Research Excellence (I-CORE) “Circle of Light” funded by the Israel Science Foundation (Grant No. 1802/12). M.R. gratefully acknowledges the support of the European Research Council (ERC) under the European Union Horizon 2020 Research and Innovation Programme (Grant Agreement No. 678862), and the Villum Foundation. C.-E. B. gratefully acknowledges support by the Swiss National Science Foundation under Division II. GR and KS are grateful for support from the NSF through DMR-1410435, the Institute of Quantum Information and Matter, an NSF Frontier center funded by the Gordon and Betty Moore Foundation, the Packard Foundation, and from the ARO MURI W911NF-16-1-0361 ``Quantum Materials by Design with Electromagnetic Excitation" sponsored by the U.S. Army. KS is additionally grateful for support from NSF Graduate Research Fellowship Program.

\appendix
\section{Floquet-Kinetic Equations \label{sec:Derivation} }

In this section, we outline the derivation of the kinetic equations for the Floquet single-particle correlation function $F_{\beta p}^{\alpha p} = \ex{f_{\alpha p}^{\dag}(t)f_{\beta p}(t)}$, where $f_{\alpha p}^{\dag}(t)$ is the creation operator for a single particle in a Floquet state with band $\alpha$ and momentum $p$. For notational convenience, we use the label order $\alpha p$ here in the appendix instead of $k\alpha$ as used in the main text. Note that $\hbar=1$ as before.

We begin by moving to the free Floquet basis via the transformation (see main text): $c_{\nu p} = \sum_{\alpha}\qb{\nu p}{\psi_{\alpha p}(t)}f_{\alpha p}(t) = \sum_{\alpha n}e^{-i(\s E_{\alpha p}+n\Omega)t}\qb{\nu p}{\phi_{\alpha p}^{n}}f_{\alpha p}(t)$, where in the second equality we have expanded the periodic part of the Floquet state $\kett{\phi_{\alpha p}(t)}$ in terms of its harmonics. Defining $U_0(t,t')$ as the time-evolution operator from $t'$ to $t$ associated with the free part of the model, $H_0(t)$, we take note of the important property of Floquet states, $f_{\alpha p}^{\dag}(t)=U_0(t,t')f_{\alpha p}^{\dag}(t')U_0^{\dag}(t,t')$. This immediately leads to the fact that $i\frac{\d}{\d t}\p{f_{i_{1}}^{\dag}(t)...f_{i_{m}}^{\dag}(t)f_{i_{m+1}}(t)...f_{i_{m+n}}(t)}=[H_{0}(t),f_{i_{1}}^{\dag}(t)...f_{i_{m}}^{\dag}(t)f_{i_{m+1}}(t)...f_{i_{m+n}}(t)]$, where $i=(\alpha,p)$ is a compressed index used for brevity. Hence, from considering both the time derivative of the state and the time derivative of the creation/annihilation operators, we obtain
$i\frac{\d}{\d t}\ex{f_{i_{1}}^{\dag}(t)...f_{i_{m}}^{\dag}(t)f_{i_{m+1}}(t)...f_{i_{m+n}}(t)} = \ex{[f_{i_{1}}^{\dag}(t)...f_{i_{m}}^{\dag}(t)f_{i_{m+1}}(t)...f_{i_{m+n}}(t),H-H_{0}(t)]}$,
where $H = H_0(t)+H_\mathrm{int}+H_\mathrm{el-ph}$ is the full Hamiltonian of the driven many-body problem. Using the above properties, we perform the cluster expansion to second order, treating doublets at the scattering level \cite{Kira2012,seetharam2015}. The major approximation in this procedure is that we factorize higher-order correlators (``doublets") into 2-point functions (``singlets")
\begin{eqnarray}
\ex{f_{i_{1}}^{\dag}f_{i_{2}}^{\dag}f_{i_{3}}f_{i_{4}}} & \approx & F_{i_{4}}^{i_{1}}F_{i_{3}}^{i_{2}}-F_{i_{3}}^{i_{1}}F_{i_{4}}^{i_{2}}\nonumber\\
\ex{f_{j_{1}}^{\dag}f_{i_{1}}^{\dag}f_{j_{2}}f_{i_{2}}b_{\boldsymbol{q}}^{\dag}b_{\boldsymbol{q}'}} & \approx & (F_{i_{2}}^{j_{1}}F_{j_{2}}^{i_{1}}-F_{j_{2}}^{j_{1}}F_{i_{2}}^{i_{1}})\ex{b_{\boldsymbol{q}}^{\dag}b_{\boldsymbol{q}'}}.
\end{eqnarray}

Furthermore, we assume that the bosons are thermal, $\ex{b_{\boldsymbol{q}}^{\dag}b_{\boldsymbol{q}'}}  \approx  \delta_{\boldsymbol{qq}'}\s N_{\omega_{\boldsymbol{q}}}$, where $\s N_{\omega_{\boldsymbol{q}}} = (e^{\beta_\mathrm{ph} \omega_{\boldsymbol{q}}}-1)^{-1}$ is the Bose-Einstein distribution with inverse temperature $\beta_\mathrm{ph}=1/T_{\mathrm{ph}}$ (we set $k_\mathrm{B}=1$). Finally, we assume the bath interactions and electron-electron interactions are Markovian (and also drop principle-value terms). Non-Markovian effects are an interesting topic and beyond the scope of this work. At this level of approximation, one obtains the \textit{Floquet-Redfield} (FRE) equation ~\cite{seetharamThesis} which couples the kinetic equations of the off-diagonal Floquet-``polarizations" (or single-particle coherences) and the diagonal Floquet occupations. The FRE requires care in its simulation as it is explicitly time-dependent and oscillatory. 
To obtain an intuitive closed set of kinetic equations for the dominant Floquet occupations alone, we keep only the occupation terms and perform the secular approximation on the remaining explicit time-dependence to obtain the \textit{Floquet-Boltzmann} (FBE) equation
\cite{seetharam2015,Bilitewski2014,GenskeRosche2015}(note that $\omega_\mathbf{q}=\omega_\mathbf{-q}$ for the acoustic phonons used here). This kinetic equation is what one would obtain if considering a ``Floquet-Fermi-Golden-Rule'' approach where the time derivative of the occupations is given by collision integrals involving scattering of electrons with each other and with phonons:
\begin{eqnarray}
\d_{t}F_{\alpha p} & = & \s G_{\mathrm{scat},+}^{\alpha p}+\s G_{\mathrm{scat},-}^{\alpha p}+\s V_{\mathrm{scat}}^{\alpha p},
\label{eq:FBE}
\end{eqnarray}
where $\s G_{\mathrm{scat},+}^{\alpha p}$ and $\s G_{\mathrm{scat},-}^{\alpha p}$ denote the two pieces of the collision integral encoding electron-phonon scattering, $I_{\alpha p}^{\mathrm{ph}}\{F\} = \s G_{\mathrm{scat},+}^{\alpha p}+\s G_{\mathrm{scat},-}^{\alpha p}$, and $I_{\alpha p}^{\mathrm{ee}}\{F\} = \s V_{\mathrm{scat}}^{\alpha p}$ denotes the collision integral encoding electron-electron scattering. Explicitly:

\begin{widetext}

\begin{eqnarray}
\s G_{\mathrm{scat},+}^{\alpha p} & = & 2\pi\sum_{\alpha_{2}p_{2}q\boldsymbol{q}_{\perp}}\sum_{n}|G_{\alpha_{2}p_{2}q}^{\alpha p}(n)|^{2}\delta(\s E_{\alpha p}-\s E_{\alpha_{2}p_{2}}-\omega_{\boldsymbol{q}}+n\Omega)\p{F_{\alpha_{2}p_{2}}(1-F_{\alpha p})\s N_{\omega_{\boldsymbol{q}}}-F_{\alpha p}(1-F_{\alpha_{2}p_{2}})(1+\s N_{\omega_{\boldsymbol{q}}})}\\
\s G_{\mathrm{scat},-}^{\alpha p} & = & 2\pi\sum_{\alpha_{2}p_{2}q\boldsymbol{q}_{\perp}}\sum_{n}|G_{\alpha_{2}p_{2}q}^{\alpha p}(n)|^{2}\delta(\s E_{\alpha p}-\s E_{\alpha_{2}p_{2}}+\omega_{\boldsymbol{q}}+n\Omega)\p{F_{\alpha_{2}p_{2}}(1-F_{\alpha p})(1+\s N_{\omega_{\boldsymbol{q}}})-F_{\alpha p}(1-F_{\alpha_{2}p_{2}})\s N_{\omega_{\boldsymbol{q}}}}\\
\s V_{\mathrm{scat}}^{\alpha p} & = & 4\pi\sum_{\alpha_{2}\alpha_{3}\alpha_{4}}\sum_{p_{2}p_{3}p_{4}}\sum_{n}|V_{\alpha_{3}p_{3}\alpha_{4}p_{4}}^{\alpha p\alpha_{2}p_{2}}(n)|^{2}\delta(\s E_{\alpha p}+\s E_{\alpha_{2}p_{2}}-\s E_{\alpha_{3}p_{3}}-\s E_{\alpha_{4}p_{4}}+n\Omega)\nonumber\\
 &  & [(1-F_{\alpha p})(1-F_{\alpha_{2}p_{2}})F_{\alpha_{3}p_{3}}F_{\alpha_{4}p_{4}}-F_{\alpha p}F_{\alpha_{2}p_{2}}(1-F_{\alpha_{3}p_{3}})(1-F_{\alpha_{4}p_{4}})],
\end{eqnarray}
where the $\alpha$ indices denote Floquet bands and the $p$ indices denote electronic momenta. As before, $q$ denotes the phonon momentum along the direction of the system, $\s{E}_{\alpha p}$ denotes the quasienergy of Floquet band $\alpha$ and momentum $p$, and $n$ is an integer characterizing the number of drive quanta exchanged in the scattering process. Moreover, $G_{\alpha_{2} p_{2} q}^{\alpha p}(n)$ and $V_{\alpha_{3}p_{3}\alpha_{4}p_{4}}^{\alpha p\alpha_{2}p_{2}}(n)$ are the dressed matrix elements which arise from changing basis to the Floquet states, given by


\begin{eqnarray}
G_{\alpha_{2}p_{2}q}^{\alpha p}(n) & = & \sum_{m\nu\nu' l}M_{\nu p_{2}}^{\nu'pq}\delta(p-p_{2}-q+2\pi l/a)
  \qb{\phi_{\alpha p}^{n+m}}{\nu'p}\qb{\nu p_{2}}{\phi_{\alpha_{2}p_{2}}^{m}}\nonumber\\
V_{\alpha_{3}p_{3}\alpha_{4}p_{4}}^{\alpha p\alpha_{2}p_{2}}(n) & = & \sum_{\nu_{1}\nu_{2}\nu_{3}\nu_{4}}\sum_{n'mm'}V_{\nu_{3}p_{3}\nu_{4}p_{4}}^{\nu_{1}p\nu_{2}p_{2}}\qb{\phi_{\alpha p}^{n-n'+m+m'}}{\nu_{1}p}  \qb{\phi_{\alpha_{2}p_{2}}^{n'}}{\nu_{2}p_{2}}\qb{\nu_{3}p_{3}}{\phi_{\alpha_{3}p_{3}}^{m}}\qb{\nu_{4}p_{4}}{\phi_{\alpha_{4}p_{4}}^{m'}},
 \label{eq:dressedElements}
\end{eqnarray}
where $\nu,\nu'$ are the undriven band indices, and where we have assumed that the bare coupling in Eq.~(\ref{eq:elbosHam}) does not depend on $\boldsymbol{q}_{\perp}$, for simplicity. Crystal-momentum conservation is explicitly shown with $l\in\mathbb{Z}$. Since we are interested in the case of a 3D bosonic bath coupled to the 1D system, we integrate out the bath degrees of freedom transverse to the system and replace the energy/momentum conservation in the FBE with a partial density of states (pDOS) defined as $\sum_{q\boldsymbol{q}_{\perp}}[\cdot] = \sum_{q}\int d\omega\rho(q,\omega)[\cdot] $. Evaluating the pDOS for the kinematic constraints yields the replacement rule (with momentum conservation up to reciprocol lattice vectors implicity assumed) $\sum_{q\boldsymbol{q}_{\perp}} \delta(p-p_2-q) \delta(\omega-\Delta E)\rightarrow\rho(p-p_2,\Delta E)$ in the FBE. For the case of linear dispersion, $\omega_{\boldsymbol{q}}=C|\boldsymbol{q}|$, the pDOS is given by 
\begin{eqnarray}
\rho(q,\omega) & = & \begin{cases}
\frac{2A_{\perp}}{(2\pi)^{2}}\frac{\pi\omega}{C^{2}} & C\sqrt{q^{2}}\leq\omega<C\sqrt{q^{2}+(\frac{\pi}{a_{b}})^{2}}\\
\frac{2A_{\perp}}{(2\pi)^{2}}\frac{2\omega}{C^{2}}\p{\mathrm{sin^{-1}}(\sqrt{\frac{(\frac{\pi}{a})^{2}}{(\frac{\omega}{C})^{2}-q^{2}}})-\mathrm{sin^{-1}}(\sqrt{1-\frac{(\frac{\pi}{a})^{2}}{(\frac{\omega}{C})^{2}-q^{2}}})} & C\sqrt{q^{2}+(\frac{\pi}{a_{b}})^{2}}\leq\omega<C\sqrt{q^{2}+2(\frac{\pi}{a_{b}})^{2}},
\end{cases}
\end{eqnarray}
where $C(\sqrt{3}\pi/a_b) = \Omega_D$, where $\Omega_D$ is the Debye frequency cutoff (see main text), 
and $A_{\perp}$ is the transverse area of the bath. With these definitions, we may further define the overall electron-phonon scattering strength
%
\begin{eqnarray}
B_{\alpha_{2}p_{2}}^{\alpha p,\pm}(n) & = & |G_{\alpha_{2}p_{2},p-p_{2}}^{\alpha p}(n)|^{2}\rho(p-p_{2},\pm(\s E_{\alpha p}-\s E_{\alpha_{2}p_{2}}+n\Omega)), \nonumber\\
\label{eq:PHscatstrength}
\end{eqnarray}
to obtain
\begin{eqnarray*}
\s G_{\mathrm{scat},\pm}^{\alpha p} & = & 2\pi\sum_{\alpha_{2}p_{2}}\sum_{n}B_{\alpha_{2}p_{2}}^{\alpha p,\pm}(n)\p{F_{\alpha_{2}p_{2}}(1-F_{\alpha p})(\frac{1}{2}\mp\frac{1}{2}+\s N_{\omega^{\pm}})-F_{\alpha p}(1-F_{\alpha_{2}p_{2}})(\frac{1}{2}\pm\frac{1}{2}+\s N_{\omega^{\pm}})}
\end{eqnarray*}
\end{widetext}
where $\omega^\pm=\pm(\s E_{\alpha p}-\s E_{\alpha_{2}p_{2}}+n\Omega)$, i.e., the Bose-Einstein distribution is evaluated at the energy argument of the pDOS. The scattering strength in Eq.~(\ref{eq:PHscatstrength}) scales as $1/N$ in system size and is independent of $A_\perp$ as $G_{\nu k}^{\nu'k'}\!\!(\boldsymbol{q})\sim (1/\sqrt{NA_\perp})$ (see Ref.~\onlinecite{seetharam2015}).

The nearest-neighbor interaction considered in Eq.~(\ref{eq:int}) in the band basis is
%
\begin{eqnarray*}
H_{\mathrm{int}} & = & \sum_{k_{1}k_{2}k_{3}k_{4}}\sum_{\nu_{1}\nu_{2}\nu_{3}\nu_{4}}V_{\nu_{3}k_{3}\nu_{4}k_{4}}^{\nu_{1}k_{1}\nu_{2}k_{2}}c_{k_{1}\nu_{1}}^{\dag}c_{k_{2}\nu_{2}}^{\dag}c_{k_{3}\nu_{3}}c_{k_{4}\nu_{4}}
\end{eqnarray*}
\begin{eqnarray}
\label{eq:Vcoup}
V_{\nu_{3}k_{3}\nu_{4}k_{4}}^{\nu_{1}k_{1}\nu_{2}k_{2}} & = & \s U(1+e^{i(k_{2}-k_{3})a})R{}_{k_{1},0\nu_{1}}^{*}R{}_{k_{2},1\nu_{2}}^{*}R_{k_{3},1\nu_{3}}R_{k_{4},0\nu_{4}}\nonumber\\
 & - & \s U(1+e^{i(k_{1}-k_{3})a})R{}_{k_{1},1\nu_{1}}^{*}R{}_{k_{2},0\nu_{2}}^{*}R_{k_{3},1\nu_{3}}R_{k_{4},0\nu_{4}}\nonumber\\
 & - & \s U(1+e^{i(k_{2}-k_{4})a})R{}_{k_{1},0\nu_{1}}^{*}R{}_{k_{2},1\nu_{2}}^{*}R_{k_{3},0\nu_{3}}R_{k_{4},1\nu_{4}}\nonumber\\
 & + & \s U(1+e^{i(k_{1}-k_{4})a})R{}_{k_{1},1\nu_{1}}^{*}R{}_{k_{2},0\nu_{2}}^{*}R_{k_{3},0\nu_{3}}R_{k_{4},1\nu_{4}},\nonumber\\
\end{eqnarray}
where $\s U = \frac{V_{0}}{4N}\delta(k_{1}+k_{2}-k_{3}-k_{4}+2\pi l/a)$, $N$ denotes the number of unit cells in the system, and $R_{k,s\nu} = \qb{s k}{\nu k}$ is the rotation matrix from the sublattice to the band basis, where $s=0,1$ corresponds to sublattice $A,B$ in Eq.~(\ref{eq:quadratic}). Note the fermionic symmetries
$V_{\nu_{3}k_{3},\nu_{4}k_{4}}^{\nu_{1}k_{1},\nu_{2}k_{2}}=-V_{\nu_{3}k_{3},\nu_{4}k_{4}}^{\nu_{2}k_{2},\nu_{1}k_{1}}=-V_{\nu_{4}k_{4},\nu_{3}k_{3}}^{\nu_{1}k_{1},\nu_{2}k_{2}}=V_{\nu_{4}k_{4},\nu_{3}k_{3}}^{\nu_{2}k_{2},\nu_{1}k_{1}}$. Hermiticity requires $V_{\nu_{3}k_{3},\nu_{4}k_{4}}^{\nu_{1}k_{1},\nu_{2}k_{2}}=(V_{\nu_{1}k_{1},\nu_{2}k_{2}}^{\nu_{3}k_{3},\nu_{4}k_{4}})^{*}$ for the interaction matrix elements and $G_{\nu k}^{\nu'k'q \boldsymbol{q}_{\perp}}=(G_{\nu'k'}^{\nu k(-q)(-\boldsymbol{q}_{\perp})})^{*}$ for the electron-phonon matrix elements.

All of the collision integrals have three main ingredients: dressed matrix elements, kinematic restrictions from the delta functions containing quasienergy (and crystal-momentum conservation hidden in the matrix elements), and phase-space factors due to Fermi and Bose statistics (occupation functions). The kinematic restrictions give crucial insight into the structure of the FBE. The scattering of a Floquet-quasiparticle via the absorption or emission of a phonon and the $2\rightarrow2$ scattering of Floquet-quasiparticles both conserve quasienergy up to multiples of the drive frequency. This kinematic structure is a signature of the fact that quasienergy is itself defined modulo $\Omega$. 

To understand its implications further, let us choose a gauge and define the first Floquet zone (FFZ) as shown in Fig.~\ref{fig:process}. As in the main text, we will refer to the upper band in the FFZ as the UF band and to the lower band in the FFZ as the LF band. By selecting a gauge, we have set an energetic orientation - the UF band is of higher quasienergy (positive values) than the LF band (negative values). 

We are now in a position to discuss the scattering processes which split into two broad categories we term ``normal'' and ``Floquet-Umklapp'' (FU), with the former encoding processes that maintain the energetic orientation and the latter that do not. Normal processes are those with $n=0$ in the quasienergy delta functions, and FU processes are those with $n\ne0$. This concept is best elucidated via examples for both phonon scattering and electron-electron interactions.
Importantly, when only $n=0$ processes are present, the system maintains detailed balance and the Floquet-Fermi-Dirac solution $F_{\alpha p}=(e^{\s{E}_{\alpha p}/T_\mathrm{ph}}+1)^{-1}$ for the steady state is exact \cite{GalitskiiElesin1970,seetharam2015}; this is mathematically the same as the case of the usual undriven Boltzmann equation, with quasienergy replacing energy.

Let us first understand how to interpret the terms in the Floquet-Boltzmann equation beginning with the electron-phonon terms. On the left-hand side (LHS) of the equation, we have the time derivative of the occupation of state $\alpha p$. The terms on the right-hand side (RHS) of the equation appearing with positive sign denote an ``incoming" transition $\alpha_{2}p_{2}\rightarrow \alpha p$, which can be understood by looking at the occupation factors. The initial state $\alpha_{2}p_{2}$ must have some occupation and the final state $\alpha p$ must have empty space; hence the rate is proportional to $F_{\alpha_{2}p_{2}}(1-F_{\alpha p})$. The $\s N_{\omega_{\boldsymbol{q}}}$ factor denotes phonon absorption and the $1+\s N_{\omega_{\boldsymbol{q}}}$ factor denotes phonon emission, since at $T_\mathrm{ph}=0$, the Bose-Einstein factors vanish but the ``$1$" term still encodes a finite rate of spontaneous emission into the ``vacuum." The terms with the negative sign denote the respective Hermitian conjugate processes, i.e., the ``outgoing" processes with transition $\alpha p\rightarrow \alpha_{2}p_{2}$. 

We term the processes with $n\ne0$ as ``Floquet-Umklapp'' processes since, in analogy to Bloch theory, the scattering processes are assisted by a reciprocal lattice vector, which here is $\Omega$. In sharp contrast to the ``normal'' processes, these  processes appear to go against the energy orientation we have chosen. Consider a process with $\s{E}_{\alpha p}\geq\s{E}_{\alpha_{2}p_{2}}$, where the initial state is $\alpha_2 p_2$ and the final state is $\alpha p$. It is only possible, assuming the appropriate energy phonon exists, to satisfy this condition in two ways: in the phonon-absorption term ($\s G_{\mathrm{scat},+}^{\alpha p}$) with $n=0$ (the normal process discussed earlier), and in the phonon-emission term ($\s G_{\mathrm{scat},-}^{\mathrm{\alpha p}}$) with $n<0$. 
The latter FU process shows that it is possible to have a transition from a lower quasienergy state, $\alpha_{2}p_{2}$, to a higher quasienergy state, $\alpha p$, via emission of a phonon. 

More generally, choosing a gauge, i.e., an energetic orientation, means to specify a preferred frame to view the Floquet bands that reside on a torus. Normal processes are those that obey kinematic intuition in the chosen frame. In contrast, FU processes are those that wrap around the torus. Choosing a different gauge corresponds to choosing a different frame, and processes that are called normal and FU in one frame will correspondingly switch roles in the other. From this discussion, it is clear that with the phonons, energetic restrictions on $\Omega_D$ with respect to the gap between the bands ($\Delta_A$) and the gap at the zone edge ($\Delta_B$) can selectively populate one or both of the bands. In fact, it is perhaps better to select the frame based on which band is preferentially populated, declaring that to be the LF band.

Let us turn our attention to the interaction term $\s V_{\mathrm{scat}}^{\alpha p}$. We can still segregate $n=0$ terms as normal processes and $n\ne0$ terms as FU processes. The normal processes just encode the usual $2\rightarrow2$ scattering obeying quasienergy conservation in the given frame (including Auger I processes). Since these processes do not change the total quasienergy, they only contribute to the spread of total quasienergy through the system. 
In contrast, the FU processes are still $2\rightarrow2$ scattering but with exchange of drive quanta, and, hence, are the source of energy non-conservation (when only the energies of the electrons are taken into account). There are two classes of FU scattering: The Floquet-Auger I (FA-I) processes are those in which two particles start in the same Floquet band, and only one particle switches Floquet bands with an exchange of a drive quantum. Floquet-Auger II (FA-II) processes are those in which two particles start in the same Floquet band, and both switch to the other. This is only possible with the exchange of a drive quantum (see Fig. \ref{fig:process}). Altogether, the energy absorption and the spread of quasienergy through the system via normal and FU processes are the mechanisms of heating in driven weakly-interacting systems.

\begin{figure*}
\includegraphics[width=\textwidth]{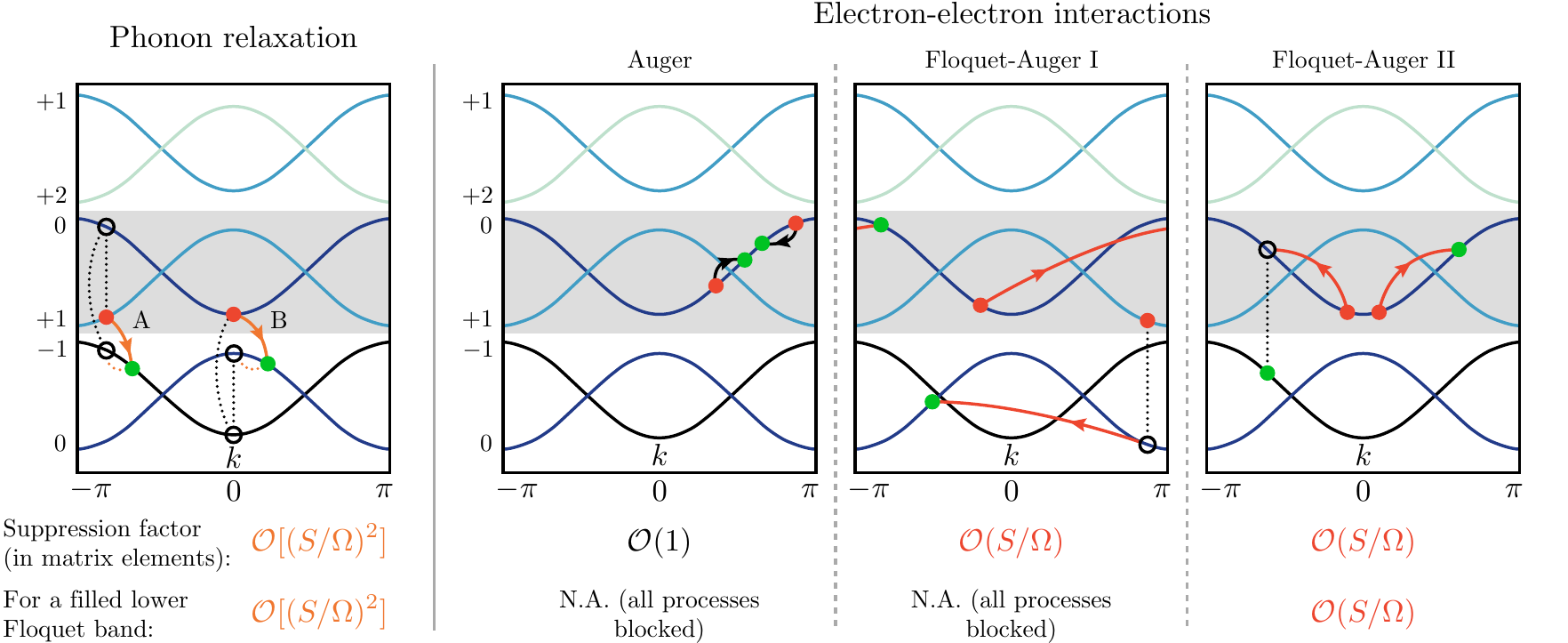}
 \caption{Dominant types of scattering processes that change the excitation density, 
classified according to their origin: phonon relaxation or electron-electron interactions (which may be of type Auger I, Floquet-Auger I, or Floquet-Auger II, described in the main text). Recall that Floquet-Auger processes of type I (II) change the number of excitations by one (two) particle(s), while absorbing energy from the drive. The energy bands shown here are copies of the bands of the non-driven system (dark blue) shifted by $m \Omega$, i.e., by integer multiples of the drive frequency. Bands are labeled by $m$, and shown in different colors for distinct $m$. The Floquet states are obtained using perturbation theory in $(S/\Omega)$ as superpositions of these harmonics. 
Here we choose our basis of Floquet states so that they have dominant harmonic components in the Floquet zone (energy window $\Omega$) highlighted in grey. Scattering processes can be decomposed into transitions between Floquet harmonics (initial/final states denoted by red/green dots), and we only illustrate the dominant ones involving leading-order harmonics. Transitions between Floquet states must conserve momentum and energy, up to an integer multiple $n \Omega$ (and up to some phonon momentum and energy, for phonon-mediated processes). Normal processes are characterized by $n=0$ (black arrows) and Floquet-Umklapp (FU) processes are characterized by $n \neq 0$ (red and orange arrows). The dotted lines indicate the drive-induced virtual transitions involved in a process, with each virtual transition bringing an additional power of the small parameter $S/\Omega$. The suppression factors of individual processes are indicated below each panel. When the lower Floquet band is filled, Auger I and Floquet-Auger I processes are absent. Note that the ``B'' phonon relaxation process can be $\mathcal{O}(1)$ if the phonon matrix elements $G_{\nu k}^{\nu'k'}(\boldsymbol{q})$ allow interband (off-diagonal in $\nu,\nu'$) transitions. An analogous scenario exists, for example, in the case of radiative recombination.}
\label{fig:scalingEl}
\end{figure*}

The last remaining ingredients of the FBE are the dressed matrix elements. The key effect of the dressing, for weak driving, is in suppressing the strength of high-$n$ scattering processes, or in other words, those that involve the exchange of many drive quanta. This comes directly from consideration of the Floquet-band matrix elements in the undriven band basis. The chosen FFZ is primarily made from the undriven conduction band and a single drive quantum shifted undriven valence band. The higher harmonic content of the FFZ states have less weight as they are detuned significantly in energy. The rates of scattering processes may strongly depend on $n$. 
See Fig.~\ref{fig:scalingEl} for more detailed information about the scaling of the dressed matrix elements in Eq.~(\ref{eq:dressedElements}) as functions of $n$.


\subsection{Effective Dynamics with Bosonic Reservoir}

Here we derive the effective model presented in the main text. 
Setting $F_{\alpha p} = n_\alpha$ to be uniform ($p$-independent) and using the half-filling condition $\sum_\alpha n_\alpha = 1$, we obtain Eq.~(\ref{eq:effDyn}) of the main text (reproduced here for convenience):
%
\begin{eqnarray}
\dot{n} & = & W_{\mathrm{in}}^{\mathrm{ph}}(1-n)^{2}-W_{\mathrm{out}}^{\mathrm{ph}}n^{2}\nonumber\\
 & + & W_{31}^{\mathrm{ee}}((1-n)^{3}n-(1-n)n^{3})+W_{22}^{\mathrm{ee}}((1-n)^{4}-n^{4}),\nonumber
\end{eqnarray}
with the following definitions. The electron-phonon rates are [using $\alpha=+$ for Eq.~(\ref{eq:effDyn})]: 
%
\begin{eqnarray}
W_{\mathrm{in}}^{\mathrm{ph},\alpha} & = & 2\pi N\sum_{m}\s B_{\bar{\alpha}}^{\alpha+}(m)\s N_{m}+2\pi N\sum_{m}\s B_{\bar{\alpha}}^{\alpha-}(m)(1+\s N_{m})\nonumber\\
W_{\mathrm{out}}^{\mathrm{ph},\alpha} & = & 2\pi N\sum_{m}(1+\s N_{m})\s B_{\bar{\alpha}}^{\alpha+}(m)+2\pi N\sum_{m}\s B_{\bar{\alpha}}^{\alpha-}(m)\s N_{m}.\nonumber\\
\label{eq:avgRatesPH}
\end{eqnarray}
where $\bar{\alpha}$ is the opposite of $\alpha$, i.e., for $\alpha=\pm$, $\bar{\alpha}=\mp$. In the equation above, we use
\begin{eqnarray}
\s B_{\alpha_{2}}^{\alpha\pm}(m) & = & \frac{1}{N^{2}}\sum_{pp_{2}}B_{\alpha_{2}p_{2}}^{\alpha p,\pm}(m),
\end{eqnarray}
which averages the scattering strengths [Eq.~(\ref{eq:PHscatstrength})] over all initial and final momenta. The average rates in Eq.~(\ref{eq:avgRatesPH}), while appearing proportional to system size $N$, are in fact \emph{intensive} as the scattering strength scales as $1/N$ [see discussion below Eq.~(\ref{eq:PHscatstrength})].
%
In addition, we neglect the momentum/energy dependence of the Bose-Einstein distribution factors $\s N_{\omega_{\boldsymbol{q}}}$ in the rates, using
\begin{eqnarray}
\s N_{m} & = & \begin{cases}
\s N_{\Delta_{A}} & m=0\\
\s N_{\Delta_{B}} & |m|=1.
\end{cases}
\end{eqnarray}
In this way, for $m=0$ (normal) processes we set the energies in all Bose-Einstein factors equal to $\Delta_A$, while for $|m| = 1$ (FU) processes we set the energies in the Bose-Einstein factors equal to $\Delta_B$.

Similarly, the transition rates arising from electron-electron interactions are
%
\begin{eqnarray}
W_{22}^{\mathrm{ee}} & = & 4\pi N^{3}V_{D}^{2},\\
W_{31}^{\mathrm{ee}} & = & 8\pi N^{3}V_{F}^{2},
\label{eq:avgRatesEE}
\end{eqnarray}
where the momentum-averaged electron-electron scattering strengths are 
\begin{eqnarray}
\s S_{\alpha_{3}\alpha_{4}}^{\alpha\alpha_{2}} & = & \frac{1}{N^{4}}\sum_{pp_{2}p_{3}p_{4}}\sum_{n}|V_{\alpha_{3}p_{3}\alpha_{4}p_{4}}^{\alpha p\alpha_{2}p_{2}}(n)|^{2}\nonumber\\
 &  & \delta(\s E_{\alpha p}+\s E_{\alpha_{2}p_{2}}-\s E_{\alpha_{3}p_{3}}-\s E_{\alpha_{4}p_{4}}+n\Omega),
\label{eq:avgEEstrength}
\end{eqnarray}
%
\begin{eqnarray}
\mkern-18mu \p{\begin{array}{cccc}
V_{1}^{2} & V_{F}^{2} & V_{F}^{2} & V_{2}^{2}\\
V_{F}^{2} & V_{D}^{2} & V_{2}^{2} & V_{F}^{2}\\
V_{F}^{2} & V_{2}^{2} & V_{D}^{2} & V_{F}^{2}\\
V_{2}^{2} & V_{F}^{2} & V_{F}^{2} & V_{1}^{2}
\end{array}} & \equiv & \p{\begin{array}{cccc}
\s S_{00}^{00} & \s S_{01}^{00} & \s S_{00}^{01} & \s S_{01}^{01}\\
\s S_{10}^{00} & \s S_{11}^{00} & \s S_{10}^{01} & \s S_{11}^{01}\\
\s S_{00}^{10} & \s S_{01}^{10} & \s S_{00}^{11} & \s S_{01}^{11}\\
\s S_{10}^{10} & \s S_{11}^{10} & \s S_{10}^{11} & \s S_{11}^{11}
\end{array}} .
\label{eq:Vmat}
\end{eqnarray}
where for notational similarity, we define the $V_1,V_2,V_F,V_D$ variables squared as equal to the various scattering strengths. Note that $\s S_{\alpha_{3}\alpha_{4}}^{\alpha\alpha_{2}}\sim 1/N^3$ by Eq.~(\ref{eq:Vcoup}). This is because $\s U^2$ provides a factor of $1/N^2$ (ignoring momentum conservation) and the momentum delta function eliminates one of the momentum sums in Eq.~(\ref{eq:avgEEstrength}). The remaining three sums over momenta provide a factor of $N^3$ and so we achieve the result that electron-electron scattering strengths scale with system size as $1/N^3$. Therefore, the electron-electron transition rates in Eq.~(\ref{eq:avgRatesEE}) are \emph{intensive}.

The matrix structure in Eq.~(\ref{eq:Vmat}) directly follows from fermionic antisymmetry, hermiticity, and particle-hole/chiral symmetry. Using the matrix $\mathcal{S}$ we assign a single parameter for each type of scattering process to characterize its average strength; 
FA-II processes have strength $V_D$, the sum of Auger and FA-I processes together have strength $V_F$, fully intraband scattering  has strength $V_1$, and interband scattering that conserves band density has strength $V_2$. As expected, only $V_F,V_D$ contribute to the effective dynamics in Eq.~(\ref{eq:effDyn}), since they are the only process types that change the band density.

\section{Fermionic Reservoir \label{sec:Lead}}

In this section we modify the effective model to include the effects of coupling to a (non-driven) Fermi reservoir.
We take a site-dependent tunnel coupling $\Gamma_{l}^{s x}$ for a lead electron $l$ tunneling into a (real-space, sublattice) system state $(x,s)$.
The Hamiltonians for the lead and the lead-system coupling are given by:
%
\begin{eqnarray}
H_{\mathrm{lead}} & = & \sum_{l}\epsilon_{l}d_{l}^{\dag}d_{l},\\
H_{\mathrm{el-lead}} & = & \sum_{axl}\Gamma_{l}^{sx}(c_{xs}^{\dag}d_{l}+d_{l}^{\dag}c_{xs})\nonumber\\
 & = & \sum_{\nu kl}\Gamma_{l}^{\nu k}c_{k\nu}^{\dag}d_{l}+h.c.,
\end{eqnarray}
where $\Gamma_{l}^{\nu k} = 1/\sqrt{N}\sum_{sx}e^{-ikx}\Gamma_{l}^{sx}R_{k,\nu s}^{\dag}$ is the tunnel coupling in the band basis. The results are derived in the same fashion as in Appendix \ref{sec:Derivation} and here we just present the main results. The corresponding collision integral that enters the FBE [Eq.~(\ref{eq:FBE})] is given by:
\begin{widetext}
%
\begin{eqnarray}
\label{eq:Rlead}\s R^{\alpha p}_{\mathrm{scat}} & = & 2\pi\sum_{l}\sum_{n}|\Gamma_{l}^{\alpha p}(n)|^{2}\delta(\s E_{\alpha p}-\epsilon_{l}+n\Omega)[(1-F_{\alpha p})D_{l}-F_{\alpha p}(1-D_{l})],
\end{eqnarray}
\end{widetext}
where $\Gamma_{l}^{\alpha k}(n) = \sum_{\nu}\Gamma_{l}^{\nu k}\qb{\phi_{\alpha k}^{n}}{\nu k}$ is the dressed lead coupling and $D_l$ is the Fermi-Dirac distribution of the lead with chemical potential $\mu_{\mathrm{res}}$ and temperature $T_{\mathrm{res}}$. 

Equation (\ref{eq:Rlead}) encodes the tunneling of a lead electron $l$ into Floquet state $(\alpha,p)$ with strength $|\Gamma_{l}^{\alpha p}(n)|^{2}$ if the lead-electron energy and the system quasienergy are matched up to $n\Omega$. Both normal and FU tunneling processes may be present based on the number of drive quanta exchanged. Detailed analysis in the context of lead engineering has been carried out in Refs.~\onlinecite{seetharam2015} and \onlinecite{IadecolaChamon2015-2}. Averaging the collision integral in Eq.~(\ref{eq:Rlead}) over all momenta, we obtain the system-lead coupling contributions to the effective model:
\begin{eqnarray}
\nonumber \dot{n}_{\alpha}  &=&  (1-n_{\alpha})\Gamma_{\mathrm{in}}^{\alpha}- n_{\alpha}\Gamma_{\mathrm{out}}^{\alpha}\\
\nonumber \Gamma_{\mathrm{in}}^{\alpha}  &=&  2\pi\sum_{l}D_{l}\bar{\Gamma}_{l}^{\alpha}\\ 
\Gamma_{\mathrm{out}}^{\alpha}  &=& 2\pi\sum_{l}(1-D_{l})\bar{\Gamma}_{l}^{\alpha}.
\end{eqnarray}
Here we have defined the momentum-averaged tunneling rates $\bar{\Gamma}_{l}^{\alpha} = 1/N\sum_{p}\sum_{n}|\Gamma_{l}^{\alpha p}(n)|^{2}\delta(\s E_{\alpha p}-\epsilon_{l}+n\Omega)$. Note that the lack of particle conservation in the presence of a lead requires one to separately consider each band density $n_\alpha$. The ``in" rates increase the number of particles in a given band, while the ``out'' rates empty those states.

We can gain intuition for the effect of the lead terms by considering the case where $\mu_\mathrm{res}=0$, $T_\mathrm{res}=0$, i.e., a zero-temperature lead with chemical potential set in the center of the gap between the two Floquet bands. 
In this case, $D_l=\Theta(\mu_\mathrm{res}-\epsilon_l)=\Theta(-\epsilon_l)$, where $\Theta$ is the Heaviside step function. Focusing on the LF band, we find that $\Gamma_{\mathrm{in}}^{-}\neq 0$ for $n\leq0$ and $\Gamma_{\mathrm{out}}^{-}\ne0$ for $n>0$, since $\s{E}_{p-}<0$ for all $p$. This means that tunneling into the LF band can occur as a normal or as an FU process (by absorbing one or more photons from the drive). In contrast, tunneling out of the LF band can only occur as an FU process. With no further restrictions, the reservoir will generically heat the system since FU tunneling processes involving exchange of drive quanta are present. However, if one considers a ``filtered'' lead with a bandwidth less than $\Omega$ (still centered between the bands), then FU processes are kinematically forbidden and $\Gamma_\mathrm{out}^-=0$. Hence, particles can only tunnel into the LF band. In the UF band the situation is reversed; with a filtered lead we have $\Gamma_\mathrm{in}^+=0$ such that particles may only tunnel out of the UF band. Therefore, a filtered lead pushes the system toward the Floquet insulator steady state with $n_-=1$ and $n_+=0$. This scenario has been analyzed in detail in Ref.~\onlinecite{seetharam2015}.


The full effective model in the presence of both bosonic and fermionic reservoirs is explicitly given by: 
\begin{widetext}
\begin{eqnarray}
\dot{n}_{\alpha} & = & (1-n_{\alpha})\Gamma_{\mathrm{in}}^{\alpha}-n_{\alpha}\Gamma_{\mathrm{out}}^{\alpha}+W_{\mathrm{in}}^{\mathrm{ph},\alpha}n_{\bar{\alpha}}(1-n_{\alpha})-W_{\mathrm{out}}^{\mathrm{ph},\alpha}n_{\alpha}(1-n_{\bar{\alpha}})\nonumber\\
\mkern-18mu & + & W_{22}^{\mathrm{ee}}\p{(1-n_{\alpha})^{2}n_{\bar{\alpha}}^{2}-n_{\alpha}^{2}(1-n_{\bar{\alpha}})^{2}}+\frac{1}{2}W_{31}^{\mathrm{ee}}\p{(1-n_{\alpha})^{2}n_{\alpha}n_{\bar{\alpha}}-(1-n_{\alpha})(1-n_{\bar{\alpha}})n_{\alpha}^{2}+(1-n_{\alpha})(1-n_{\bar{\alpha}})n_{\bar{\alpha}}^{2}-(1-n_{\bar{\alpha}})^{2}n_{\alpha}n_{\bar{\alpha}}}.\nonumber\\
 \label{eq:FullEffDyn}
\end{eqnarray}
\end{widetext}
Equation~(\ref{eq:FullEffDyn}) simplifies to Eq.~(\ref{eq:effDyn}) in the case of half-filling and no fermionic reservoir. 

\section{Simulation Details \label{sec:Simulation} }

We use the electronic hopping parameter $J_1/J_0=-0.425$, drive parameters $S/J_0=0.5$, $\Omega/J_0=1.65$, and normalize all length scales by the electronic lattice spacing $a$ (i.e. set $a=1$), corresponding to gaps in the Floquet spectrum of $\Delta_A/J_0=0.2$, $\Delta_B/J_0=0.45$. The phonons have velocity $C=(\Omega_D/\sqrt{3})(a_b/\pi)$ (with $a_b=a$), with spectral cutoff (bandwidth) $\Omega_D$, and temperature $T=\Delta_A/10$. The interaction strength is $V_0/J_0=0.25$. 

The delta function enforcing quasienergy conservation in electron-electron collisions, appearing in the collision integral $\s{V}^\mathrm{\alpha p}_\mathrm{scat}$, is approximated on the finite-size system with a Gaussian of finite support:
\begin{eqnarray}
\delta(\Delta\s E) & \approx & \begin{cases}
\frac{Z(r)}{\sqrt{2\pi\varepsilon^{2}}}e^{-\frac{(\Delta\s E)^{2}}{2\varepsilon^{2}}}, & |\Delta\s E|\leq r\varepsilon\\
0, & \mathrm{o.w.},
\end{cases}
\end{eqnarray}
where for the standard deviation we take $\varepsilon=\mathrm{max}_k(\s{E}_{\alpha,k+2\pi/(Na)}-\s{E}_{\alpha,k})$ is the maximum adjacent quasienergy level spacing in a single Floquet band, $r=1.5$ denotes the number of deviations to include in the finite support, and $Z(r=1.5)=1.154$ is the normalization constant ensuring that the truncated Gaussian function integrates to unity. By allowing finite support in quasienergy to the delta function, one is, in a rough sense, adding a linewidth to the quasienergy states. One must check that these linewidths are smaller than the Floquet gaps, as otherwise the approximation introduces unphysical interband transitions not appearing in the FBE (e.g., one particle stays in the same state and the other is directly excited across the gap at the same momentum). 
We check that the truncated Gaussian does not allow such anomalous transitions across the Floquet gaps by ensuring that $r\varepsilon<\Delta_A,\Delta_B$. 
In the FBE simulations ($N=20$), we scan the amplitude $G_0$ of the phonon bath and the cutoff $\Omega_D$, and perform numerical integration of the FBE until reaching a steady state for each choice of $G_0$ and $\Omega_D$.

\bibliographystyle{apsrev}
\bibliography{main-refs}

\end{document}